# Characterization of Mood and Emotion Regulation in Females with PMS/PMDD Using Near-Infrared Spectroscopy to Assess Prefrontal Cerebral Blood Flow and the Mood Questionnaire

Makiko Aoki*
Faculty of Health Science and Nursing
Juntendo University
Shizuoka, Japan
m.aoki.tt@juntendo.ac.jp

Masato Suzuki
Graduate School of Science
University of Hyogo
Hyogo, Japan
suzuki@sci.u-hyogo.ac.jp
https://orcid.org/0000-0003-2040-3029

Satoshi Suzuki
Department of Clinical Engineering
Kanagawa Institute of Technology
Kanagawa, Japan
suzuki@cet.kanagawa-it.ac.jp

Kosuke Oiwa
Department of Information and Management Systems Engineering
Nagaoka University of Technology
Niigata, Japan
oiwa@vos.nagaokaut.ac.jp
https://orcid.org/0000-0002-7508-171X

Yoshitaka Maeda
Medical Simulation Centre
Jichi Medical University
Tochigi, Japan
y-maeda@jichi.ac.jp
https://orcid.org/0000-0003-4895-2820

Hisayo Okayama
Faculty of Medicine
University of Tsukuba
Ibaraki, Japan
okayama@md.tsukuba.ac.jp
https://orcid.org/0000-0003-1253-9874

*Abstract*—Many sexually mature women experience premenstrual syndrome (PMS) or premenstrual dysphoric mood disorder (PMDD). This discomfort, being recurrent and cyclical, significantly impacts women's quality of life. Current approaches for managing PMS/PMDD rely on daily mental condition recording, which many discontinue due to its impracticality. Hence, there's a critical need for a simple, objective method to monitor mental symptoms. One of the principal symptoms of PMDD is a dysfunction in emotional regulation, which has been demonstrated through brain-function imaging measurements to involve hyperactivity in the amygdala and a decrease in functionality in the prefrontal cortex (PFC). However, most research has been focused on PMDD, leaving a gap in understanding of PMS. Near-infrared spectroscopy (NIRS) measures brain activity by spectroscopically determining the amount of hemoglobin in the blood vessels. The NIRS device is compact and imposes minimal physical and psychological stress on participants. This study aimed to characterize the emotional regulation function in PMS. We measured brain activity in the PFC region using NIRS when participants were presented with emotion-inducing pictures. Furthermore, moods highly associated with emotions were assessed through questionnaires. Forty-six participants were categorized into non-PMS, PMS, and PMDD groups based on the gynecologist's diagnosis. The mood of participants was collected by a self-administered questionnaire (POMS2) and NIRS was used to obtain cerebral blood flow. POMS2 scores revealed higher negative mood and lower positive mood in the follicular phase for the PMS group, while the PMDD group exhibited heightened negative mood during the luteal phase. NIRS results showed reduced emotional expression in the PMS group during both phases, while no significant differences were observed in the PMDD group compared to non-PMS. It was found that there are differences in the distribution of mood during the luteal and follicular phase and in cerebral blood flow responses to emotional stimuli between PMS and PMDD. These findings suggest the potential for providing individuals with awareness of PMS or PMDD through scores on the POMS2 and NIRS measurements.

## I. Introduction

The menstrual cycle is a 28-35 day cycle, with premenstrual syndrome (PMS) often occurring during the luteal phase, which is marked by psychological and physical discomforts. When these psychiatric symptoms progress further a more severe form of PMS, known as premenstrual dysphoric disorder (PMDD), is diagnosed [1]. The economic loss in Japan due to physical and mental disorders associated with the menstrual cycle is estimated to be JPY 638 billion [2]. An effective approach to alleviating or coping with PMS or PMDD is awareness and acceptance of PMS/PMDD symptoms [3]. However, it requires daily recording of one's mental condition over a long time, and many females discontinue the practice [4]. To alleviate the unique premenstrual symptoms in women, it is necessary to establish a method that allows for the easy, simple, objective, and continuous self-awareness of mental symptoms such as PMS and PMDD.

Emotional dysregulation in females with PMS/PMDD is one of the most frequently occurring mental symptoms. For instance, females with PMS have been reported to exhibit a heightened sensitivity to anxiety [5], and increased resilience to negative emotions and stress [6], [7]. Functional magnetic resonance imaging (fMRI) in females with PMDD has shown hyperactivity in the amygdala [8], which governs emotional control, and a decrease in activity in the prefrontal cortex (PFC) region upon presentation of emotion-inducing pictures [9], [10]. These previous studies indicate that it is possible to objectively measure the mental symptoms of PMDD. However, most previous studies have primarily focused on participants with PMDD, with few reports on brain activity during emotional stimuli in participants with PMS. In this study, to elucidate the characteristics of emotional regulation functions in participants with PMS, the brain function of the PFC region was objectively measured through brain imaging techniques when they were presented with emotion-inducing pictures. Additionally, the emotions and moods of the PMS participants were subjectively assessed using questionnaires. Using the results from both subjective and objective assessments, the emotional regulation functions in PMS participants were characterized.

*Corresponding Author
This work was supported by JSPS KAKENHI Grant Numbers 21K12794 to HO and 22K12950 to MA.

Furthermore, this study utilized near-infrared spectroscopy (NIRS) as a method for measuring brain function. NIRS measures the concentration of hemoglobin in the blood through spectroscopic techniques, allowing for a more compact measurement device with less restraint on the participants compared to fMRI. Simultaneous measurements of cerebral blood flow in the PFC region of participants performing cognitive tasks using both NIRS and fMRI have reported similarities in signals [11]. Additionally, we have reported a decrease in cerebral blood flow in the PFC of PMS subjects during cognitive tasks using NIRS [12]. In this study, to clarify the characteristics of emotional and mood regulation functions in PMS, a mood of participants with PMS were evaluated by mood questionnaires, and measurements of hemoglobin in the PFC region were conducted using NIRS, which allows for easy measurement of activation in the PFC region and imposes low restraint on the participants.

## II. EXPERIMENTAL METHODS

### A. Requirements for participants in this study

Healthy female university students aged 20-25 years were recruited as participants. Participants were selected based on the following selection criteria:

- BMI between 18.0 and 25.0.
- Self-Depression Scale (SDS) less than 60 points.
- Not taking low-dose contraceptives for more than 3 months.
- No smoking habit.
- No history of psychiatric disorders and no ongoing medications for the treatment of psychiatric disorders.
- No experience of pregnancy or childbirth.
- No experience of gynecological problems and disorders.

### B. Experimental Procedure

Participants who agreed to informed consent were given a self-administered questionnaire (age, height, weight, medical history, age at menarche, menstrual cycle menstrual periods), SDS [13], [14], PMDD scale [15], and basal body thermometer (MC-652-LC, OMRON Corp, Kyoto, Japan). They were asked to measure and record their basal body temperature for two menstrual cycles. The basal body temperature charts and PMDD scales were interpreted by a gynecologist, and PMS was diagnosed. In the second or third week of the luteal phase and follicular phase after starting the basal body temperature recording, participants were asked to fill out the mood scale (the profile of mood states 2) [16], and cerebral blood flow was measurements by NIRS. The NIRS probe (OEG-SpO$_2$, Spectratech Inc. Tokyo, Japan) was attached to the participant's prefrontal area. Emotional stimulus images were displayed on an LCD monitor (23.8 in, 1,920 pix × 1,080 pix) positioned 55 cm in front of the participants, and cerebral blood flow was measured.

### C. Methods of presenting emotional stimulus images

Images eliciting positive emotions were selected from the International Affective Pictures System (IAPS) [17]. The average emotional valence of these 20 images ranged from 5.7 to 6.3, and the average arousal level ranged from 4.5 to 5.0. Emotional valence and arousal level are indicators used to quantify emotions. Emotional valence indicates pleasantness or unpleasantness on a scale of 9 (pleasant: 9, unpleasant: 1), while arousal measures the extent to which an image is stimulating or calming on a 9-point scale (high arousal: 9, low arousal: 1). For eliciting negative emotions, images from the Disgust-Related-Images (DIRTI) [18] were used, selecting those with a valence of 1.1 to 2.0 and an arousal level of 4.5 to 5.4.

### D. Evaluation of cerebral blood flow using NIRS

The NIRS probe (469 × 60 mm) consisted of channels (Ch) 1-16, with Ch1 placed on the right temporal region and Ch16 on the left temporal region. As most of the participants in this study had a small facial structure, hair interfered with the placement of the temporal probes, making it difficult to securely attach all channels to the forehead area of the participants. Therefore, the average NIRS signal obtained from Ch4 to Ch7 was used as the NIRS data for the right brain, and the average values from Ch10 to Ch13 were used for the left brain. For the measurement of oxygenated and deoxygenated hemoglobin in the blood vessels, near-infrared light with wavelengths of 770 nm or 840 nm was emitted from the light-emitting parts of the NIRS probe toward the forehead, and scattered and reflected light was detected by the photodetector within the NIRS probe. The sampling interval for detection was 0.65 seconds. The average NIRS signals, representing a concentration of oxygenated hemoglobin (Oxy-Hb), from 5 seconds before to just before the presentation of the colored image was used as the baseline. The subtraction between the NIRS signal measured at each time point and the baseline was calculated and a sum of these from 10 seconds to 25 seconds in Fig. 1 was obtained. This sum was termed the integral value, which represents the change in Oxy-Hb concentration following the image presentation. The integral value was obtained for each stimulus set, and the mean of the integral values from 10 stimulus sets was calculated to serve as the representative value.

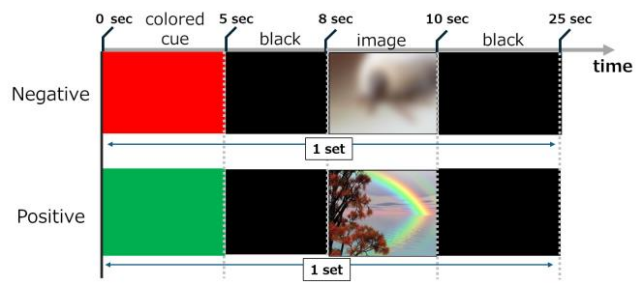

Fig. 1. Schematic diagram illustrating the anticipation of emotional stimuli.

TABLE I.    MEAN AGE, MEAN BMI, AND MEAN MENSTRUAL CYCLE FOR PMDD, PMS, AND NON-PMS PARTICIPANTS.

|  | PMDD group (5 participants) | | | PMS group (16 participants) | | | Non-PMS group (25 participants) | | | $p^*$ |
|---|---|---|---|---|---|---|---|---|---|---|
|  | *Mean* | *SD* | *SE* | *Mean* | *SD* | *SE* | *Mean* | *SD* | *SE* |  |
| Age, year | 21.4 | 0.9 | 0.3 | 20.8 | 0.9 | 0.3 | 21.1 | 0.8 | 0.1 | .25 |
| BMI, kg m$^{-2}$ | 20.4 | 1.7 | 0.6 | 21.3 | 3.2 | 0.9 | 20.5 | 2.8 | 0.5 | .88 |
| Menstrual cycle, day | 31.7 | 3.2 | 1.2 | 30.4 | 3.0 | 0.8 | 30.2 | 3.8 | 0.6 | .43 |

a. Kruskal-Wallis test. BMI: body mass index.

### E. Data Analysis

Emotional valence, arousal values, POMS2 scores, and NIRS signal integrals were compared between two of the three groups: non-PMS, PMS, and PMDD by the Wilcoxon rank-sum test (Bonferroni correction).

### F. Ethical Consideration

This study was conducted after obtaining approval from the Medical Ethics Committee of Medicine and Health Sciences in the University of Tsukuba Faculty of (Approval No. 1650-1).

## III. RESULTS AND DISCUSSION

### A. Basic properties of the participants

Among 46 participants, 25 participants were categorized as non-PMS, 16 participants as PMS, and 5 participants as PMDD. Table 1 summarizes the mean age, mean BMI, and mean menstrual cycle length for the non-PMS, PMS, and PMDD groups, respectively. No significant differences were observed among the three groups in terms of age, BMI, and menstrual cycle length, confirming that the groups were well-balanced.

### B. Valence and arousal levels among the three groups during the anticipation of emotional stimuli

Figure 2 represents valence and arousal in the participants of non-PMS, PMS, or PMDD during the anticipation of emotional stimuli. In the non-PMS group, participants evoked pleasant feelings toward positive images and unpleasant towards negative images, regardless of whether it was the follicular or luteal phase. This suggests that participants in the non-PMS group consistently processed emotions appropriately, irrespective of their menstrual cycle phase. Conversely, in the PMS/PMDD groups, the emotional valence for positive images was significantly lower than that of the non-PMS group during the follicular phase ($p$=.002), while valence for negative images was significantly higher ($p$=.04). This indicates that participants in the PMS/PMDD groups experienced less emotional evocation for positive stimuli and less discomfort for negative stimuli. These findings suggest that the PMS/PMDD group tends to be less emotionally expressive in the follicular phase. However, a previous study comparing emotional valence between PMDD and non-PMDD reported no significant differences in both positive and negative emotional valence [19], which is different from the present results. One reason for this discrepancy might be that the average SDS scores for the participants in the present study were above 40 for both the luteal and follicular phases. These relatively high SDS scores could mean that the participants were mildly depressed and may have had slightly higher levels of depression than the population in the previous study. Individuals with depressive symptoms have been reported to be insensitive to emotional stimuli [20], [21]. Our findings suggest that participants with PMS/PMDD tend to be less expressive in emotional stimulation during the follicular phase compared to the luteal phase. On the other hand, no significant differences in arousal were observed among the groups.

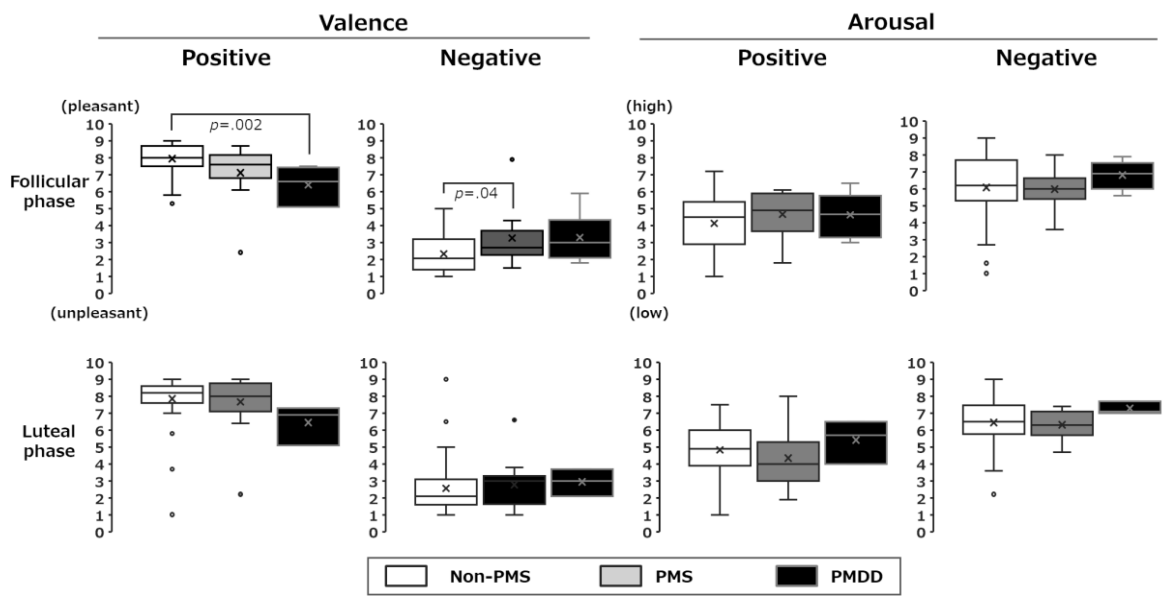

Fig. 2. The emotional valence and arousal levels of participants from the non-PMS, PMS, and PMDD groups when presented with images eliciting either positive or negative emotions.

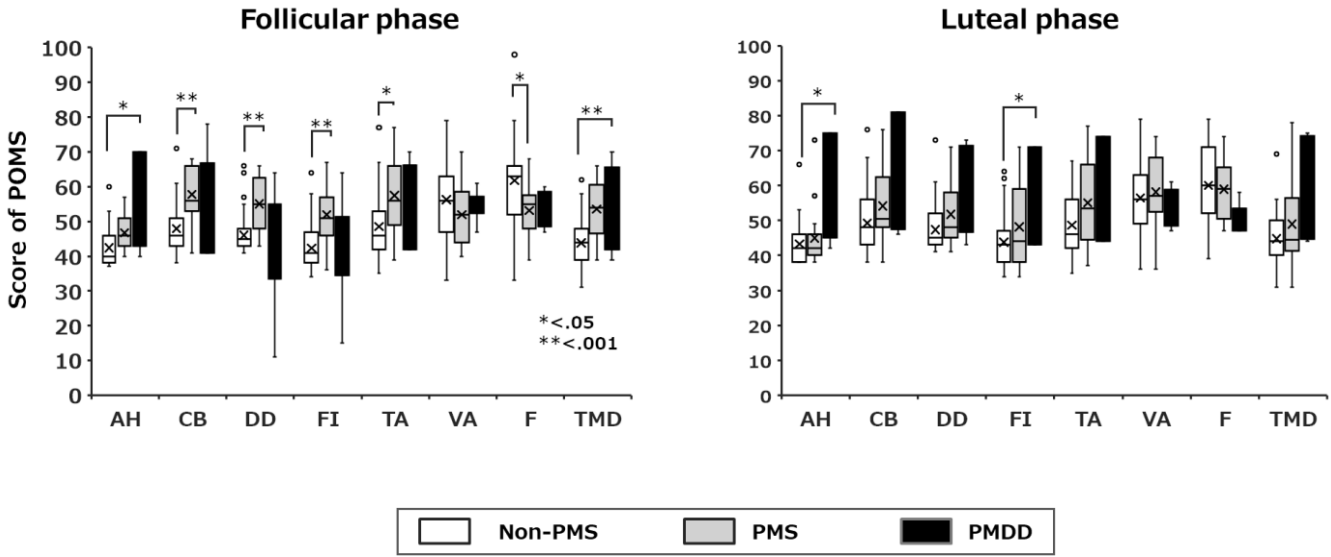

Fig. 3. Scores for seven sub-scales of POMS2 for participants in non-PMS, PMS, and PMDD during the follicular and luteal phase. The data are shown as the mean and the error bars are standard errors. AH, Anger-hostility; CB, Confusion-bewilderment; DD, Depression-dejection; FI, Fatigue-inertia; TA, Tension-anxiety; VA, Vigor-activity; F, Friendliness; TMD, Total mood disturbance.

*C. Comparison of POMS2 Scores Among the Three Groups*

The results of scores of POMS2 for participants in non-PMS, PMS, and PMDD groups are shown in Figure 3. In the follicular phase, participants in PMS showed significantly higher scores compared to non-PMS in anger-hostility (AH), confusion-bewilderment (CB), depression-dejection (DD), fatigue-inertia (FI), tension-anxiety (TA), and total mood disturbance (TMD) ($p<.05$, respectively), and significantly lower in Friendliness (F). These results indicate that in the follicular phase, females with PMS experience higher negative mood and lower positive mood compared to the non-PMS group, suggesting that PMS symptoms may originate from the follicular phase. Previous research reported that females with PMS can be classified according to their mood in the follicular phase [22]. Although PMS is typically considered to be symptomatic in the luteal phase [23], our findings suggest that PMS-derived symptoms may be expressed even in the follicular phase. Conversely, participants in PMDD showed significantly higher scores compared to non-PMS in AH and FI during the luteal phase ($p<.05$, respectively). This indicates that females with PMDD may experience heightened negative mood and discomfort during the luteal phase. PMDD is specifically defined as a condition where psychological symptoms appear especially in the luteal phase [24], and our findings support that participants with PMDD exhibit symptoms during this phase. These results suggest that PMS may cause mood alterations starting from the follicular phase, while PMDD may cause mood disturbances in the luteal phase.

*D. Integrated values of Oxy-Hb change during the anticipation of emotional stimuli.*

The results of integral values obtained from ch4-ch7 during stimulation with positive emotional anticipation images among non-PMS, PMS, and PMDD groups are shown in Figure 4. The Integral Value of the non-PMS group during the follicular phase tended to be higher compared to the PMS group ($p=.08$). In the integral values at the luteal phase, the PMS group was significantly lower than the non-PMS. This indicates lower brain activation in the follicular and luteal phases for the PMS group, suggesting a decrease in emotional expression within this group. Considering the results of the scores of POMS2 for the PMS group, it appears that females with PMS are associated with a decrease in emotional expression and mood alterations starting from the follicular phase. Thus, mood and emotional expression during the follicular phase could potentially serve as biomarkers for characterizing PMS symptoms. Conversely, no significant differences were observed in the integrated values of Oxy-Hb changes between the PMDD group and the non-PMS group. This suggests that PMS may not simply represent a more severe form of PMDD, but rather, it should be considered as a distinct condition.

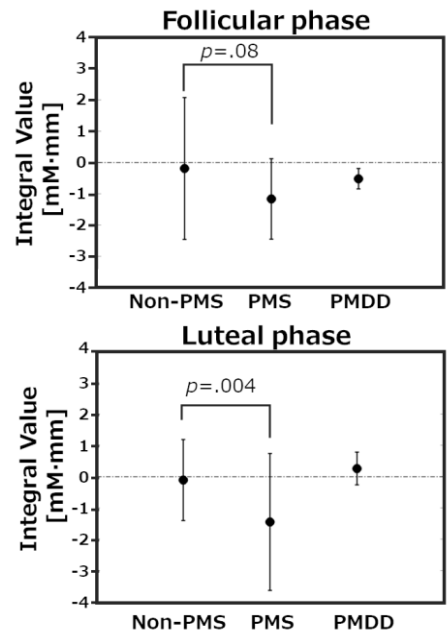

Fig. 4. The integral values obtained from ch4-ch7 of the NIRS probe of non-PMS, PMS, and PMDD participants during the anticipation of positive emotional stimuli. The data are shown as the mean of 10 positive stimulus sets and the error bars are standard deviations.

## IV. Conclusion

In this study, we aimed to elucidate the characteristics of mood and emotional regulation of females with PMS. The results revealed that participants with PMS exhibited a decrease in positive mood, an increase in negative mood during the follicular phase, and a reduction in Oxy-Hb in the PFC when presented with images evoking positive emotions. Conversely, participants with PMDD showed a significant increase in negative mood during the luteal phase, with Oxy-Hb responses similar to those of non-PMS participants. Thus, the questionnaire to assess mood and the measurement of cerebral blood flow during anticipation of emotional stimuli produced different results for the PMS and PMDD participants. These findings suggest that integrating mood indicators with NIRS measurements of cerebral blood flow can enable the differentiation of PMS and PMDD, which has been challenging to distinguish previously. However, this study has a limitation in its small sample size, especially the number of PMDD participants, making the reproducibility of these results limited. Future research plans to increase the sample size to further clarify the characteristics of mood and emotional regulation functions in individuals with PMS and PMDD.

## Acknowledgment

M. A. and M. S. performed experiments. M. A., S. S., K. O., and Y. M. analyzed experimental data. M. A., M. S., and H. O. conceived the project. M. A. and M. S. wrote the paper.

## References

[1] L. A. Futterman and A. J. Rapkin, "Diagnosis of premenstrual disorders," *J. Reprod. Med.*, vol. 51, no. 4 Suppl, pp. 349–358, Apr. 2006.

[2] M. Song and H. Kanaoka, "Effectiveness of mobile application for menstrual management of working women in Japan: randomized controlled trial and medical economic evaluation," *J. Med. Econ.*, vol. 21, no. 11, pp. 1131–1138, Nov. 2018, doi: 10.1080/13696998.2018.1515082.

[3] S. Nayman, D. T. Konstantinow, I. F. Schricker, I. Reinhard, and C. Kuehner, "Associations of premenstrual symptoms with daily rumination and perceived stress and the moderating effects of mindfulness facets on symptom cyclicity in premenstrual syndrome," *Arch. Womens Ment. Health*, vol. 26, no. 2, pp. 167–176, Apr. 2023, doi: 10.1007/s00737-023-01304-5.

[4] T. Takeda, K. Yoshimi, S. Kai, and F. Inoue, "The Japanese Version of the Daily Record of Severity of Problems for Premenstrual Symptoms: Reliability and Validity Among the General Japanese Population," *Int. J. Womens Health*, vol. 16, pp. 299–308, Feb. 2024, doi: 10.2147/IJWH.S450300.

[5] J. Craner, S. Sigmon, A. Martinson, and M. McGillicuddy, "Perceptions of health and somatic sensations in women reporting premenstrual syndrome and premenstrual dysphoric disorder," *J. Nerv. Ment. Dis.*, vol. 201, no. 9, pp. 780–785, Sep. 2013, doi: 10.1097/NMD.0b013e3182a213f1.

[6] M. Kleinstäuber, K. Schmelzer, B. Ditzen, G. Andersson, W. Hiller, and C. Weise, "Psychosocial Profile of Women with Premenstrual Syndrome and Healthy Controls: A Comparative Study," *Int. J. Behav. Med.*, vol. 23, no. 6, pp. 752–763, Dec. 2016, doi: 10.1007/s12529-016-9564-9.

[7] K. Watanabe and T. Shirakawa, "Characteristics of Perceived Stress and Salivary Levels of Secretory Immunoglobulin A and Cortisol in Japanese Women With Premenstrual Syndrome," *Nurs. Midwifery Stud.*, vol. 4, no. 2, p. e24795, Jun. 2015, doi: 10.17795/nmsjournal24795.

[8] M. Gingnell, A. Morell, E. Bannbers, J. Wikström, and I. Sundström Poromaa, "Menstrual cycle effects on amygdala reactivity to emotional stimulation in premenstrual dysphoric disorder," *Horm. Behav.*, vol. 62, no. 4, pp. 400–406, Sep. 2012, doi: 10.1016/j.yhbeh.2012.07.005.

[9] S. Toffoletto, R. Lanzenberger, M. Gingnell, I. Sundström-Poromaa, and E. Comasco, "Emotional and cognitive functional imaging of estrogen and progesterone effects in the female human brain: a systematic review," *Psychoneuroendocrinology*, vol. 50, pp. 28–52, Dec. 2014, doi: 10.1016/j.psyneuen.2014.07.025.

[10] N. Petersen, D. G. Ghahremani, A. J. Rapkin, S. M. Berman, L. Liang, and E. D. London, "Brain activation during emotion regulation in women with premenstrual dysphoric disorder," *Psychol. Med.*, vol. 48, no. 11, pp. 1795–1802, Aug. 2018, doi: 10.1017/S0033291717003270.

[11] H. Doi, S. Nishitani, and K. Shinohara, "NIRS as a tool for assaying emotional function in the prefrontal cortex," *Front. Hum. Neurosci.*, vol. 7, Nov. 2013, doi: 10.3389/fnhum.2013.00770.

[12] M. Aoki, M. Suzuki, S. Suzuki, H. Takao, and H. Okayama, "Cognitive function evaluation in premenstrual syndrome during the follicular and luteal phases using near-infrared spectroscopy," *Compr. Psychoneuroendocrinology*, vol. 10, p. 100117, May 2022, doi: 10.1016/j.cpnec.2022.100117.

[13] W. W. K. ZUNG, "A Self-Rating Depression Scale," *Arch. Gen. Psychiatry*, vol. 12, no. 1, pp. 63–70, Jan. 1965, doi: 10.1001/archpsyc.1965.01720310065008.

[14] Kazuhiko Fukuda and Shigeo Kobayashi, "Reseach for Self-rating depression scale," *Psychiatr. Neurol. Jpn.*, vol. 75, no. 10, pp. 673–679, 1973.

[15] Y. Miyaoka, Y. Akimoto, K. Ueda, and T. Kamo, "The reliability and validity of the newly developed PMDD scale," *J. Jpn. Soc. Psychosom. Obstet. Gynecol.*, vol. 14, no. 2, pp. 194–201, 2009, doi: 10.18977/jspog.14.2_194.

[16] Heuchert, J. P. and& McNair, D. M., "Profile of Mood States, 2nd Edition: POMS 2." Multi-Health Systems Inc., 2012.

[17] Lang, P. J, Bradley, M. M, and Cuthbert, B. N., "International Affective Picture System (IAPS): Instruction manual and affective ratings, Technical Report A-8." The Center for Research in Psychophysiology, University of Florida., 2008.

[18] A. Haberkamp, J. A. Glombiewski, F. Schmidt, and A. Barke, "The DIsgust-RelaTed-Images (DIRTI) database: Validation of a novel standardized set of disgust pictures," *Behav. Res. Ther.*, vol. 89, pp. 86–94, Feb. 2017, doi: 10.1016/j.brat.2016.11.010.

[19] M. Gingnell, E. Bannbers, J. Wikström, M. Fredrikson, and I. Sundström-Poromaa, "Premenstrual dysphoric disorder and prefrontal reactivity during anticipation of emotional stimuli," *Eur. Neuropsychopharmacol.*, vol. 23, no. 11, pp. 1474–1483, Nov. 2013, doi: 10.1016/j.euroneuro.2013.08.002.

[20] M. Feeser et al., "Context insensitivity during positive and negative emotional expectancy in depression assessed with functional magnetic resonance imaging," *Psychiatry Res.*, vol. 212, no. 1, pp. 28–35, Apr. 2013, doi: 10.1016/j.pscychresns.2012.11.010.

[21] L. M. Bylsma, "Emotion context insensitivity in depression: Toward an integrated and contextualized approach," *Psychophysiology*, vol. 58, no. 2, p. e13715, Feb. 2021, doi: 10.1111/psyp.13715.

[22] M. Aoki, M. Hoshino, M. Suzuki, and H. Okayama, "Distinction between women with premenstrual syndrome or premenstrual dysphoric disorder and healthy women based on clustering Profile of Mood States 2nd Edition scores in the follicular phase," *J. Nurs. Sci. Eng.*, vol. 9, pp. 108–116, 2022, doi: 10.24462/jnse.9.0_108.

[23] ACOG Clinical Practice Guideline No. 7, "Management of Premenstrual Syndrome," *BJOG Int. J. Obstet. Gynaecol.*, vol. 124, no. 3, pp. e73–e105, 2017, doi: 10.1111/1471-0528.14260.

[24] T. Beddig, I. Reinhard, and C. Kuehner, "Stress, mood, and cortisol during daily life in women with Premenstrual Dysphoric Disorder (PMDD)," *Psychoneuroendocrinology*, vol. 109, p. 104372, Nov. 2019, doi: 10.1016/j.psyneuen.2019.104372.